\documentclass[aps,prb,twocolumn,superscriptaddress,10pt,nofootinbib]{revtex4-2}
\setcitestyle{numbers,square}
\pdfoutput=1

\usepackage[utf8]{inputenc}
\usepackage[T1]{fontenc}
\usepackage[english]{babel} 
\usepackage{amsmath, amsfonts, amssymb}
\usepackage[locale=US, separate-uncertainty=true, range-units=single, list-units=single, retain-unity-mantissa=false]{siunitx}
\usepackage{xspace}
\usepackage{nicefrac}
\usepackage{bm}
\usepackage{bbm}
\usepackage{graphicx}
\usepackage[dvipsnames]{xcolor}
\usepackage{hyperref}
\hypersetup{
	colorlinks = true,
	linkcolor = blue,
	citecolor = blue,
	filecolor = blue,
	urlcolor = blue
}

\usepackage{prettyref}%
\newcommand{\pref}[1]{\prettyref{#1}}%
\newrefformat{fig}{Fig.~\ref{#1}}%
\newrefformat{tab}{Table~\ref{#1}}%
\newrefformat{sec}{Sec.~\ref{#1}}%
\newrefformat{app}{App.~\ref{#1}}%
\newrefformat{eq}{Eq.~(\ref{#1})}%

\newcommand{\iu}{\ensuremath{\mathrm{i}}} %i nicht kursiv durch \iu
\newcommand{\eg}{\ensuremath{{e_g}}\xspace}
\newcommand{\ttg}{\ensuremath{{t_{2g}}}\xspace}
\newcommand{\BMRO}{Ba$_2$MgReO$_6$\xspace}
\newcommand{\BNOO}{Ba$_2$NaOsO$_6$\xspace}
\newcommand{\ReOoct}{ReO$_6$\xspace}
\newcommand{\curlyU}{\ensuremath{{\mathcal{U}}}\xspace}
\newcommand{\curlyJ}{\mathcal{J}}
\newcommand{\Jeff}{\ensuremath{{J_\mathrm{eff}}}\xspace}
\newcommand{\bra}[1]{\langle #1 |}
\newcommand{\ket}[1]{| #1 \rangle}
\newcommand{\braket}[2]{\langle #1 | #2 \rangle}
\newcommand{\Fmtm}{$Fm\bar3m$\xspace}
\newcommand{\Pftmnm}{$P4_2/mnm$\xspace}
\newcommand{\onlinearia}{Ref.~\onlinecite{mansouri_tehrani_untangling_2021}\xspace}

\DeclareSIUnit\angstrom{\text {Å}}

\begin{document}

\title{Probing the Mott-insulating behavior of Ba$_2$MgReO$_6$ with DFT+DMFT}
\author{Maximilian E. Merkel}
\affiliation{Materials Theory, ETH Z\"u{}rich, Wolfgang-Pauli-Strasse 27, 8093 Z\"u{}rich, Switzerland}
\author{Aria Mansouri Tehrani}
\affiliation{Materials Theory, ETH Z\"u{}rich, Wolfgang-Pauli-Strasse 27, 8093 Z\"u{}rich, Switzerland}
\affiliation{Research Laboratory of
Electronics, Massachusetts Institute of Technology, Cambridge, MA 02139, USA}
\author{Claude Ederer}
\email{claude.ederer@mat.ethz.ch}
\affiliation{Materials Theory, ETH Z\"u{}rich, Wolfgang-Pauli-Strasse 27, 8093 Z\"u{}rich, Switzerland}

\date{\today}

\begin{abstract}
We investigate the interplay of spin-orbit coupling, electronic correlations, and lattice distortions in the $5d^1$ double perovskite Ba$_2$MgReO$_6$. Combining density-functional theory (DFT) and dynamical mean-field theory (DMFT), we establish the Mott-insulating character of Ba$_2$MgReO$_6$ in both its cubic and tetragonal paramagnetic phases. Despite substantial spin-orbit coupling, its impact on the formation of the insulating state is minimal, consistent with theoretical expectations for $d^1$ systems. 
We further characterize the electronic properties of the cubic and tetragonal phases by analyzing spectral functions and local occupations in terms of multipole moments centered on the Re sites. Our results confirm the presence of ferroically ordered $z^2$ quadrupoles in addition to the antiferroic $x^2-y^2$-type order.
We compare two equivalent but complementary descriptions in terms of either effective Re-\ttg frontier orbitals or more localized atomic-like Re-d and O-p orbitals. The former maps directly on a physically intuitive picture in terms of nominal $d^1$ Re cations, while the latter explicitly demonstrates the role of hybridization with the ligands in the spin-orbit splitting and the formation of the charge quadrupoles around the Re sites.
Finally, we compare our DFT+DMFT results with a previous DFT+$U$ study of the tetragonal paramagnetic state. We find good qualitative agreement for the dominant charge quadrupoles, but also notable differences in the corresponding spectral functions, underscoring the need for more comparative studies between these two methods.
\end{abstract}

\maketitle

\section{Introduction}\label{sec:introduction}

In materials containing heavy elements with open $4d$, $5d$, $4f$, or $5f$ shells, the interplay of spin-orbit coupling (SOC) and electronic correlations can lead to the emergence of various exotic phases, such as quantum spin liquids, Weyl semimetals, topological insulators, superconductors, or states with magnetic octupolar order and quenched magnetic moments~\cite{hanawa_superconductivity_2001, jackeli_mott_2009, chen_exotic_2010, witczak-krempa_correlated_2014, pourovskii_ferro-octupolar_2021}.
SOC also affects the appearance of Mott-insulating states and orbital order \cite{chen_exotic_2010, bersuker_vibronic_1989, khomskii_transition_2014}. For example, for Sr$_2$IrO$_4$, it has been shown that SOC is crucial for the formation of the insulating state by reducing the orbital degeneracy~\cite{shimura_structure_1995, kim_phase-sensitive_2009, pesin_mott_2010, sato_spin-orbit-induced_2015}.
However, the effect of SOC strongly depends on the nominal filling of the electronic shell, and in general, the Mott-insulating state can be both stabilized or destabilized~\cite{kim_j_2017, triebl_spin-orbit_2018, chikano_multipolar_2021}.
Similarly, in cases where an electronic degeneracy can be lifted by a lattice distortion, SOC can either compete or cooperate with this distortion and the corresponding orbital order~\cite{bersuker_vibronic_1989, khomskii_transition_2014, khomskii_orbital_2021, streltsov_interplay_2022}.

\begin{figure}
    \centering
    \includegraphics[width=1\linewidth]{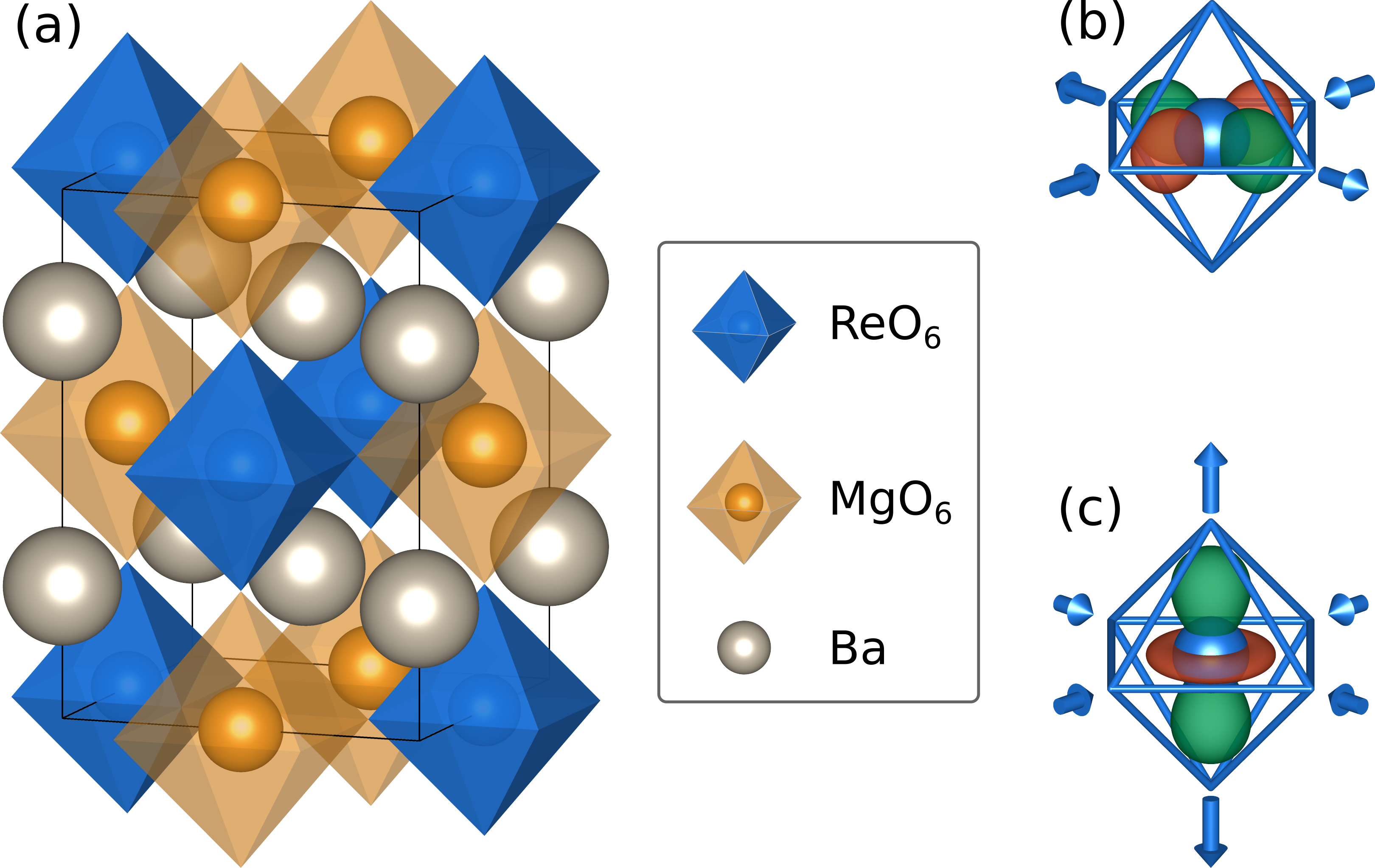}
    \caption{(a) Experimental, cubic \Fmtm crystal structure of \BMRO \cite{hirai_detection_2020}. In the tetragonal \Pftmnm phase, the \ReOoct octahedra deform with (b) rhombic $Q_2$ and (c) tetragonal $Q_3$ Jahn-Teller distortions that couple to charge quadrupoles, indicated by green regions with charge excess and red regions with charge depletion. }
    \label{fig:crystal_struct}
\end{figure}

An attractive class of materials to systematically study the interplay between SOC and electronic correlations are the double perovskites, $A_2BB'X_6$~\cite{vasala_a2bbo6_2015}, with closed-shell ions on the $A$ and $B$ sites and a heavy $4d$ or $5d$ transition-metal cation on the $B'$ site. The $B$ and $B'$ sites are arranged in a 3D-checkerboard pattern (see \pref{fig:crystal_struct}). Thus the relatively large spacing between the $B'X_6$ octahedra leads to fairly narrow transition-metal $d$ bands, such that a Mott-insulating state can be formed even for a moderately strong local electron-electron interaction~\cite{chen_exotic_2010, witczak-krempa_correlated_2014, martins_coulomb_2017, takayama_spinorbit-entangled_2021}.

Specifically, for a $d^1$ configuration of the $B'$ cation, SOC is expected to weakly stabilize the Mott-insulating state~\cite{kim_j_2017, triebl_spin-orbit_2018} and to reduce the tendency towards Jahn-Teller distortions, but without completely suppressing it~\cite{streltsov_jahn-teller_2020, khomskii_orbital_2021}.
Examples of the corresponding family of $5d^1$ double perovskites that have been studied recently are rhenates~\cite{bramnik_preparation_2003, hirai_possible_2021, ishikawa_phase_2021, frontini_spin-orbit-lattice_2023, da_cruz_pinha_barbosa_impact_2022}, \BNOO \cite{erickson_ferromagnetism_2007, lu_magnetism_2017, willa_phase_2019, fiore_mosca_interplay_2021, da_cruz_pinha_barbosa_impact_2022, fiore_mosca_mott_2023, celiberti_spin-orbital_2023}, or the more ionic Cs$_2$TaCl$_6$ \cite{ishikawa_ordering_2019, mansouri_tehrani_charge_2023}.

Here, we focus on the $5d^1$ double perovskite \BMRO \cite{longo_magnetic_1961, bramnik_preparation_2003, marjerrison_cubic_2016, hirai_successive_2019, hirai_detection_2020, mansouri_tehrani_untangling_2021, pasztorova_experimental_2023, soh_spectroscopic_2023},
which exhibits two symmetry-lowering phase transitions.
Upon cooling, the material first transitions from cubic \Fmtm symmetry to tetragonal \Pftmnm symmetry at \SI{33}{K}. This transition involves a $Q_2$-type Jahn-Teller distortion of the \ReOoct octahedra in the $xy$ plane [see \pref{fig:crystal_struct}(b)], with the sign of the distortion alternating between neighboring $xy$ planes along the $z$ direction, and an additional $Q_3$-type distortion, elongating all \ReOoct octahedra along the $z$ direction [see \pref{fig:crystal_struct}(c)] \cite{van_vleck_jahnteller_1938, hirai_detection_2020}.
At the N{\'e}el temperature of \SI{18}{K}, \BMRO then develops a canted antiferromagnetic order~\cite{hirai_successive_2019}. 
Furthermore, \BMRO is insulating at all measured temperatures, and an Arrhenius fit to electrical resistivity measurements indicates an activation energy of \SI{.17}{eV} \cite{hirai_successive_2019}.

In this paper, we use density-functional theory (DFT) plus dynamical mean-field theory (DMFT) to establish the Mott-insulating character of both the cubic and tetragonal paramagnetic phases of \BMRO. Our calculations suggest that \BMRO can be classified as a regular Mott insulator, where the insulating character is due to a small band width compared to the Coulomb repulsion $\curlyU$, which we calculate using the constrained random-phase approximation (cRPA), whereas the SOC only plays a minor role.

We then discuss the spectral properties and charge distribution of paramagnetic \BMRO in terms of local occupations and multipole moments \cite{bultmark_multipole_2009} centered on the Re sites.
Thereby, we compare two different sets of Wannier functions used to represent the Re $d$ states. First, a basis of so-called frontier orbitals, constructed only from the low-energy bands immediately around the Fermi level, and second, a more atomic-orbital-like basis of Re-$d$ and O-$p$ orbitals obtained for a larger energy window.
While the frontier basis is used in our DFT+DMFT calculations and directly maps on a nominal $d^1$ description of the SOC on the Re cations, similar to the model-based studies in Refs.~\onlinecite{kim_j_2017, triebl_spin-orbit_2018, iwahara_vibronic_2023}, the atomic-like basis reveals that both Re and O states experience the SOC and contribute to the orbital polarization in the tetragonal phase. 

The atomic-like basis also allows for a direct comparison to DFT+$U$ calculations for \BMRO from \onlinearia, where the paramagnetic structure is emulated using a large supercell with random magnetic moments on the Re atoms and zero overall magnetic moment~\cite{alling_effect_2010, trimarchi_polymorphous_2018}.
We find reasonably good agreement between the two methods for the local quadrupole moments on the Re sites, whereas the density of states obtained from the DFT+$U$ calculations differs notably from the spectral function obtained within DFT+DMFT.
Even though the different representations of the local Coulomb interaction in the two approaches can potentially explain part of this difference, a more systematic comparison would be of interest in future studies.

\section{Methods}

\subsection{Spin-orbit coupling and local electron-electron interaction}

We perform DFT+DMFT calculations for \BMRO, using a low-energy description of the material with one electron in the Re-$d$ shell. The crystal field in the \ReOoct octahedra is much stronger than the SOC so that the higher-lying \eg orbitals are nominally empty and the electron resides in the \ttg orbitals further split by the SOC.

We first perform calculations with SOC included in the DFT exchange-correlation potential and obtain the corresponding Kohn-Sham band structure.
To then systematically vary the strength of the SOC in our DMFT calculations, we compare these Kohn-Sham bands to the ones obtained when excluding SOC from the exchange-correlation potential and instead adding the local Hamiltonian \cite{pavarini_ldadmft_2018}
\begin{align}
    H_\mathrm{SOC} &= \lambda \sum_{mm'} \sum_{\sigma\sigma'} \bra{m\sigma} \bm l \cdot \bm s \ket{m'\sigma'} c_{m\sigma}^\dagger c_{m'\sigma'} \nonumber\\
    &= \frac{\iu\lambda}2 \left(
         c^\dag_{xz\uparrow} c_{xy\downarrow} + c^\dag_{xz\downarrow} c_{xy\uparrow}
         + \iu c^\dag_{yz\uparrow} c_{xy\downarrow} - \iu c^\dag_{yz\downarrow} c_{xy\uparrow} \right. \nonumber\\
         &\hphantom{=\frac{\iu\lambda}2(} \left. + c^\dag_{yz\uparrow} c_{xz\uparrow} - c^\dag_{yz\downarrow} c_{xz\downarrow}
    \right) + \text{h.c.}
    \label{eq:h_soc}
\end{align}
to the \ttg Wannier orbitals in a post-processing step, where $\lambda$ parametrizes the SOC strength.

In the presence of SOC, it is also instructive to represent our results in the eigenbasis $\ket{\Jeff, m_\Jeff}$ of the Hamiltonian in \pref{eq:h_soc}, where $\Jeff$ is the effective total angular momentum, defined as the sum of the effective orbital momentum $l=1$ of the \ttg orbitals and the spin, taking values of $\Jeff=1/2$ and $\Jeff=3/2$~\cite{sugano_multiplets_1970}.
These states are related to the \ttg orbitals by
\begin{align}
    \begin{pmatrix}
        \ket{1/2, -1/2} \\ 
        \ket{3/2, -1/2} \\
        \ket{3/2, \hphantom{-} 3/2}
    \end{pmatrix} =
    \begin{pmatrix}
        -\iu/\sqrt3 & \hphantom{-} 1/\sqrt3 & -1/\sqrt3 \\
        -\iu/\sqrt6 & \hphantom{-} 1/\sqrt6 & \hphantom{-} 2/\sqrt6 \\
        -\iu/\sqrt2 & -1/\sqrt2 & \hphantom{-} 0 
    \end{pmatrix}
    \begin{pmatrix}
        \ket{xz \uparrow} \\ \ket{yz \uparrow} \\ \ket{xy \downarrow}
    \end{pmatrix}
    \label{eq:def_jeff}
\end{align}
and equivalently for the corresponding time-reversal-symmetric states \cite{martins_coulomb_2017}).
The eigenvalues with respect to $H_\text{SOC}$ are $+\lambda$ for all $\Jeff = 1/2$ and $-\lambda/2$ for all $\Jeff=3/2$ states.

To describe the local electron-electron interactions in the DMFT calculations, we use the Kanamori Hamiltonian \cite{pavarini_charge_2018} 
\begin{align}
    H_\mathrm{int} =&
    \frac12 \sum_{m,\sigma} \curlyU n_{m\sigma} n_{m\bar\sigma}
    + \frac12 \sum_{m\neq m',\sigma} \curlyU' n_{m\sigma} n_{m'\bar\sigma} \nonumber\\
    +& \frac12 \sum_{m\neq m',\sigma} (\curlyU' - \curlyJ) n_{m\sigma} n_{m'\sigma} \nonumber\\
    +& \frac12 \sum_{m\not= m',\sigma} \curlyJ (c^\dag_{m\sigma} c^\dag_{m'\bar\sigma} c_{m\bar\sigma} c_{m'\sigma} + c^\dag_{m\sigma} c^\dag_{m\bar\sigma} c_{m'\bar\sigma} c_{m'\sigma})
    \label{eq:kanamori}
\end{align}
which is exact for real-valued \ttg-type orbitals in cubic symmetry~\cite{ribic_cubic_2014}.
The three parameters $\curlyU$, $\curlyU'$, and $\curlyJ$ describe the strength of the screened intra-orbital, inter-orbital, and Hund's interaction, respectively.

We calculate these interaction parameters using the cRPA \cite{aryasetiawan_frequency-dependent_2004, miyake_ab_2009, sasioglu_effective_2011}, but we also treat them as free parameters in our DFT+DMFT calculations to analyze the corresponding effect on the electronic properties. Thereby, the complexity is reduced by assuming approximate spherical symmetry so that there are only two free parameters with $\curlyU' = \curlyU - 2\curlyJ$ \cite{ribic_cubic_2014}.
In real materials, spherical symmetry is never exactly fulfilled. Therefore, we verify the quality of this assumption based on the interaction parameters obtained from cRPA.
We can then define averaged parameters
\begin{align}
    \curlyU_\mathrm{avg} &= (13 \curlyU + 2 \curlyU' + 4 \curlyJ)/15 \nonumber\\
    \curlyJ_\mathrm{avg} &= (\curlyU - \curlyU' + \curlyJ)/3
    \label{eq:sph_avg_kanamori}
\end{align}
that minimize the square of the deviations $\curlyU_\mathrm{avg} - \curlyU$, $(\curlyU_\mathrm{avg} - 2\curlyJ_\mathrm{avg}) - \curlyU'$ and $(\curlyU_\mathrm{avg} - 3\curlyJ_\mathrm{avg}) - (\curlyU' - \curlyJ)$, weighted by their frequency of occurrence in the density-density terms of \pref{eq:kanamori}.

\subsection{Computational details}

We perform calculations for the experimental cubic \Fmtm and tetragonal \Pftmnm structures \cite{hirai_detection_2020}, where we always use a coordinate system with the axes parallel to the Re-O bonds.
The cubic structure has a lattice constant of \SI{8.08}{\angstrom} for the \Fmtm conventional cubic cell containing four formula units, with a Re-O bond length of \SI{1.93}{\angstrom}.
In the tetragonal phase, the unit cell distorts very slightly, with $c/a = 1.001$ and $a = \SI{8.08}{\angstrom}$. Furthermore, the Re-O bond lengths change by \SIlist{-.025;.006;.019}{\angstrom} in the $x$, $z$ and $y$ direction respectively in one layer, which corresponds to $|Q_2| = \SI{.044}{\angstrom}$ and $|Q_3| = \SI{.01}{\angstrom}$ as specified in Ref.~\onlinecite{hirai_detection_2020}. 
All \ReOoct octahedra are symmetry-equivalent, with octahedra in two adjacent $xy$ layers mapped onto each other by a translation combined with a \ang{90} rotation around the $z$ axis, thereby swapping the Re-O bond lengths along $x$ and $y$.

For the DFT calculations, we use VASP (version 6.4.1)  \cite{kresse_ab_1993, kresse_efficient_1996} with the PBE exchange-correlation potential \cite{perdew_generalized_1996} including the Ba-$5s$, Ba-$5p$, Mg-$2p$ and Re-$5p$ semicore states as valence electrons.
To ensure good convergence, we use a \SI{600}{eV} plane-wave cutoff and an electronic energy convergence of \SI{e-8}{eV}.
To calculate the electronic bands for the cubic \Fmtm structure, we use the primitive cell of the face-centered cubic lattice containing one formula unit, and a $8\times8\times8$ reciprocal $k$-grid. For the DFT+DMFT calculations, we treat both the cubic and the tetragonal \Pftmnm phase using the primitive tetragonal unit cell with two formula units on a $7\times7\times5$ reciprocal grid.

When SOC is incorporated directly in the DFT exchange-correlation potential, we preserve time-reversal symmetry by initializing the magnetic moments to zero and verifying that the bands are always twofold degenerate up to a precision of \SI{1e-6}{eV} in the band structure. Without SOC, we only use a single spin channel.

We then define Wannier functions using Wannier90 (version 3.1.0) \cite{pizzi_wannier90_2020} and initial projections on atom-centered atomic orbitals with well-defined orbital character.
If applicable, we disentangle the bands with a relative convergence of the gauge-invariant spread of \num{e-9} and always perform Wannierization down to a convergence of the spread of \SI{e-9}{\angstrom^2}. The only exception is the calculation with SOC in DFT, where we use the initial projections without subsequent Wannierization since this would lead to convergence away from the orbital-spin basis to the \Jeff basis instead.

To obtain estimates of the screened interaction parameters for the local electron-electron interaction in this basis of Wannier functions, we perform cRPA calculations at zero frequency using VASP without SOC. To include enough empty bands in cRPA, we first obtain the accurate Kohn-Sham energies from exact diagonalization for 512 bands using the converged charge density.

We perform one-shot DMFT calculations (neglecting charge self-consistency) with solid\_dmft \cite{merkel_solid_dmft_2022} using the quantum Monte-Carlo solver CT-HYB \cite{seth_triqscthyb_2016} and the interface to Wannier90 as implemented in DFTTools \cite{aichhorn_triqsdfttools_2016}, all based on the TRIQS library (version 3.1.x) \cite{parcollet_triqs_2015}.
The calculations are performed at an electronic temperature of \SI{290}{K} except those discussed in \pref{app:res_scan_temp}.

We use a basis of $\ket{xz}$, $\ket{-\iu yz}$ and $\ket{\iu xy}$, for which the local Hamiltonian, consisting of the SOC Hamiltonian in \pref{eq:h_soc} and possibly the local tetragonal splitting, is real. We enforce a paramagnetic solution by averaging the impurity Green function with its time-reversed counterpart after every solver run.
For all structures, cubic and tetragonal, we keep the two \ReOoct octahedra in the unit cell symmetry-equivalent by mapping one onto the other with the \ang{90} rotation about the $z$ axis and the translation described above.
Therefore, in the following, we only show the occupations of the \ReOoct octahedron with the long bond along $x$ and report all observables per formula unit (f.u.) where applicable.
We obtain the spectral functions from the analytic continuation of the impurity Green function with MaxEnt \cite{jarrell_bayesian_1996, maxent}.

\section{Results}

\subsection{DFT calculations}
\label{sec:res_dft_crpa}

\subsubsection{Band structure}
\begin{figure}
    \centering
    \includegraphics[width=1\linewidth]{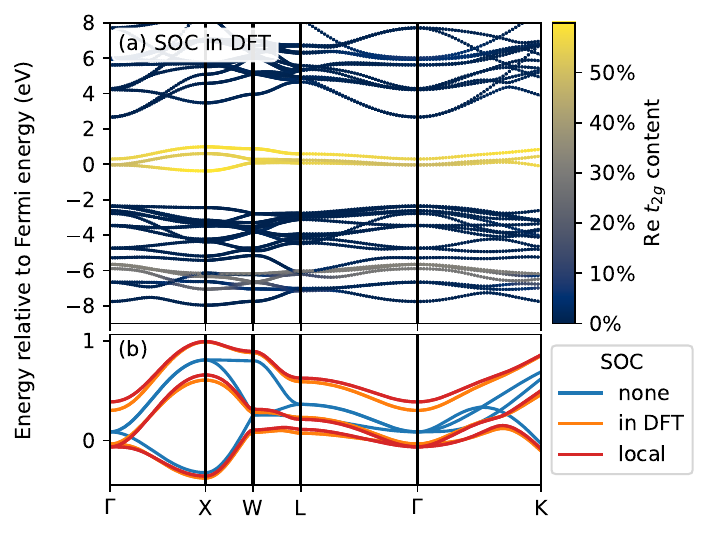}
    \caption{DFT band structure obtained for cubic \BMRO along a path of the Brillouin zone of the face-centered cubic lattice. (a) DFT bands with SOC included, and (b) bands around the Fermi energy for three different cases where the SOC is incorporated directly at the DFT level (``in DFT''), not at all (``none''), or is added as a local term in the form of \pref{eq:h_soc} to the Wannier Hamiltonian (``local'').}
    \label{fig:band_struct}
\end{figure}

We first discuss the electronic band-structure obtained within DFT for nonmagnetic, cubic \BMRO with SOC included at the DFT level. \pref{fig:band_struct}(a) shows the corresponding bands along a path in reciprocal space. The band structure is metallic and features six twofold degenerate bands with a total width of \SI{1.4}{eV} around the Fermi energy, which are separated by large gaps of around \SI{2}{eV} from bands at higher and lower energies.
Projection of these bands onto local atomic orbitals reveals their strong Re-\ttg character with an overall weight of around \SIrange{50}{60}{\%}. The second most dominant contribution comes from O-$p$ orbitals, consistent with the expected strong hybridization between Re and O. 

We construct a representation of this isolated group of bands in terms of \emph{frontier} Wannier orbitals, $\ket{\mathrm{f}_m}$, with \ttg symmetry centered at the Re cations, where $m=yz$, $xz$, and $xy$. The local part of the Hamiltonian in this Wannier basis has the form of \pref{eq:h_soc} with a strength of $\lambda = \SI{.30}{eV}$, very similar to the value of \SI{.3}{eV} reported for Os systems \cite{martins_coulomb_2017, fiore_mosca_interplay_2021}.
In addition, the Wannier Hamiltonian exhibits all intersite terms that are allowed by cubic symmetry, with the largest hopping of \SI{-0.11}{eV} between \ttg orbitals oriented along the Re-Re axis of two neighboring \ReOoct octahedra.
The small hopping amplitude, due to the large Re-Re spacing, is consistent with the small bandwidth.

Since \BMRO is experimentally known to be insulating in all its phases (including the cubic one), we investigate the effect of the local electron-electron interaction within DFT+DMFT on the electronic structure around the Fermi energy. Given the small bandwidth of the Re-\ttg dominated bands, this can potentially have a strong effect.

Within our DFT+DMFT calculations, which are presented in the following sections, we use a slightly different approach to incorporate the SOC, allowing us to treat the SOC strength as a free parameter.
We first perform the DFT calculation without SOC and again construct \ttg-like Wannier orbitals centered on the Re sites for the bands around the Fermi level. This results in a band structure with a higher degeneracy and a slightly smaller bandwidth of \SI{1.13}{eV} [blue line in \pref{fig:band_struct}(b)]. We then add the local SOC in the form of \pref{eq:h_soc} to the Wannier Hamiltonian and recalculate the corresponding bands. Indeed, by using $\lambda = \SI{.30}{eV}$ as extracted above, we obtain nearly matching bands from this approach compared to the bands obtained by incorporating SOC directly in the DFT calculation [red and orange line in \pref{fig:band_struct}(b), respectively], with a maximum deviation of \SI{.09}{eV} at the $\Gamma$ point in the highest-energy band.
Thus, even though the SOC is incorporated in the DFT calculations on an atomic level, i.e., using projections on atomic-like orbitals~\cite{steiner_calculation_2016}, its effect on the bands around the Fermi level in \BMRO can be well accounted for by \pref{eq:h_soc} applied to the Re-centered frontier orbitals only.

% ------------------------ cRPA
\subsubsection{Calculation of screened interaction parameters}

Before presenting the results of our DFT+DMFT calculations, we calculate the strength of the local electron-electron interaction in the low-energy basis with cRPA (without including SOC in the underlying DFT calculations). Thereby, the bands are separated into a target subspace, containing the set of isolated bands around the Fermi level, and a screening subspace, which consists of all other bands.
In the cubic phase, the interaction exhibits the exact Kanamori form, so that we can simply read off the Kanamori parameters from the cRPA results as $\curlyU = \SI{2.82}{eV}$, $\curlyU' = \SI{2.24}{eV}$ and $\curlyJ = \SI{0.24}{eV}$.
The deviation from spherical symmetry can be quantified by comparing $(\curlyU-\curlyU')/2$ with $\curlyJ$, which only differ by \SI{.05}{eV}, comparable to other $5d$ materials \cite{ribic_cubic_2014, fiore_mosca_mott_2023}.

Therefore, in the following, we simplify the parametrization by assuming spherical symmetry with $\curlyU' = \curlyU - 2\curlyJ$ and averaging according to \pref{eq:sph_avg_kanamori}. This results in $\curlyU = \SI{2.81}{eV}$ and $\curlyJ = \SI{.27}{eV}$.
These values are very similar to those reported for \BNOO obtained from cRPA calculations, with $\curlyU = \SI{2.9}{eV}$ and $\curlyJ = \SI{.20}{eV}$ \cite{fiore_mosca_mott_2023}, and also from an estimate based on a cluster model fitted to experimental data \cite{erickson_ferromagnetism_2007}. 
We note that, due to the large separation between the target bands and the rest, the known tendency of cRPA for overscreening is expected to be rather small~\cite{van_loon_random_2021}.
We also do not expect SOC to strongly influence the interaction parameters, since the correlated bands do not change much with or without SOC compared to the large separation from the surrounding screening bands, as shown in \pref{fig:band_struct}. This is in agreement with Refs.~\onlinecite{liu_comparative_2020, fiore_mosca_mott_2023}.

In the following DFT+DMFT calculations, we always use the value of $\curlyJ = \SI{.27}{eV}$. In \pref{sec:res_scan_u} and \pref{sec:soc-effect}, the parameter $\curlyU$ is treated as a free parameter to analyze the effect of the local interaction strength on the electronic structure, while in \pref{sec:analyis} a more detailed analysis is presented for the cRPA value of $\curlyU=\SI{2.81}{eV}$, which can be considered as realistic interaction strength for \BMRO.

\subsection{Metal-insulator transitions induced by electron-electron interaction}
\label{sec:res_scan_u}

We first explore the formation of the paramagnetic insulating state and its dependency on the Coulomb repulsion $\curlyU$.

\begin{figure}
    \centering
    \includegraphics[width=1\linewidth]{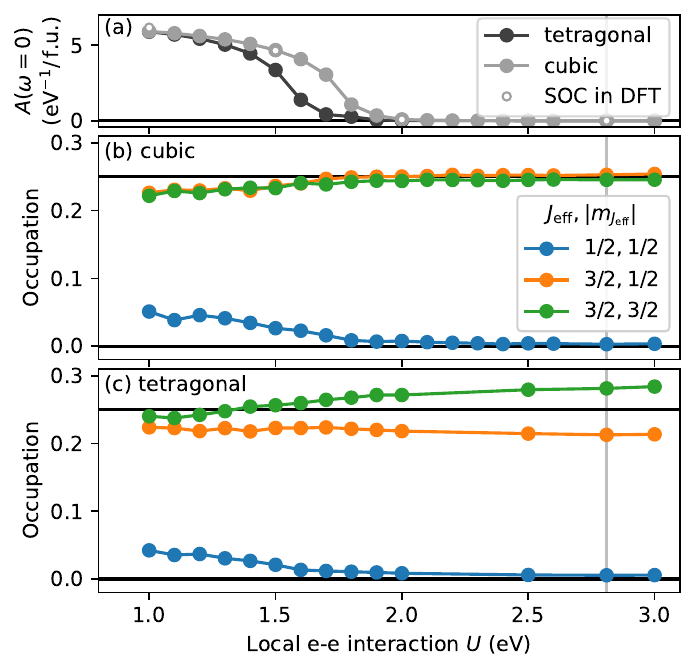}
    \caption{DFT+DMFT results for \BMRO as a function of the Coulomb interaction parameter $\curlyU$. (a) The spectral density obtained for the cubic and tetragonal structures and occupations of the \Jeff states on one Re site for the (b) cubic and (c) tetragonal phases.
    }
    \label{fig:res_scan_u}
\end{figure}

As shown by the spectral weight, $A(\omega=0)$, as a function of $\curlyU$ in \pref{fig:res_scan_u}(a), there is a metal-insulator transition for the cubic structure at a moderate $\curlyU \approx \SI{1.8}{eV}$. Since this is significantly lower than the cRPA value of $\curlyU=2.81$~eV, this shows that the electron-electron interaction is strong enough to turn cubic \BMRO insulating, also considering that cRPA tends to underestimate the screened interaction parameter, as discussed above. 
The observed critical value is in good agreement with a simple atomic-limit estimation of the metal to Mott insulator transition, where a critical $\curlyU = \SI{1.85}{eV}$ is obtained by equating the atomic-charge gap \cite{kim_j_2017} with the bandwidth of \SI{1.13}{eV} obtained without SOC.
This establishes the Mott-insulating character of \BMRO in the cubic phase.

We also obtain basically identical results from DFT+DMFT calculations where the SOC is already incorporated in the DFT calculations [gray open circles shown for some data points in \pref{fig:res_scan_u}(a)].

In all calculations for the cubic structure, the local density matrix in terms of Wannier functions is diagonal in the \Jeff basis defined in \pref{eq:def_jeff}. The evolution of these occupations as a function of $\curlyU$ is shown in \pref{fig:res_scan_u}(b).
Note that the total occupation is always 1, corresponding to the nominal Re-$t_{2g}^1$ occupation described by the low-energy frontier orbitals.
All states corresponding to the $\Jeff = 3/2$ quadruplet are always equally occupied, and the same is true for the $\Jeff = 1/2$ doublet, demonstrating that the charge density, in this case, does not spontaneously break the cubic symmetry.
In the insulating state, the electron exclusively occupies the $\Jeff = 3/2$ states, whereas in the metallic state, some occupation is transferred into the $\Jeff = 1/2$ doublet.

Next, we perform analogous DFT+DMFT calculations for the tetragonal structure of \BMRO. 
Here, the main effect of the structural distortions on the electronic structure is a local splitting of the previously degenerate \ttg Wannier functions by \SIlist{-.05;.01;.04}{eV}, while the intersite hopping terms are barely affected.

As shown in \pref{fig:res_scan_u}(a), the distortions in the tetragonal structure shift the critical $\curlyU$ for the metal-insulator transition to a slightly smaller value of $\curlyU=\SI{1.6}{eV}$ compared to the cubic case.
The distortions also affect the orbital occupations, where the occupation matrix is not diagonal in the \Jeff basis because $\Jeff$ and $m_\Jeff$ are not good quantum numbers anymore.

In \pref{fig:res_scan_u}(c), we show the diagonal elements of the occupations in the \Jeff basis for the Re site with the long-bond axis along $x$. In the insulating state, there is now a polarization within the $\Jeff = 3/2$ quadruplet, and the $\Jeff = 1/2$ states are also slightly occupied. This is due to the distortion lifting the degeneracy within the $\Jeff = 3/2$ states and coupling them to the $\Jeff = 1/2$ states.
In the metallic state, the occupations become more similar to the cubic metallic case for decreasing $\curlyU$, with almost equally occupied $\Jeff = 3/2$ states, indicating that the tetragonal crystal-field splitting is small compared to the kinetic energy.

\subsection{Effect of spin-orbit coupling strength}
\label{sec:soc-effect}

We now analyze the effect of the SOC on the metal-insulator transition in \BMRO.
In the cubic phase, SOC is known to stabilize the insulating state but has a much weaker effect for the $d^1$ case than for other fillings \cite{kim_j_2017, triebl_spin-orbit_2018}.
For example, in the atomic limit, completely turning off SOC would increase the critical $\curlyU$ from the \SI{1.85}{eV} discussed before to \SI{1.94}{eV}, which is still well below the value from cRPA.
We therefore focus on the more complex case of the tetragonal phase with its competition between tetragonal crystal-field and SOC splittings, again fixing $\curlyJ = \SI{.27}{eV}$.

\begin{figure}
    \centering
    \includegraphics[width=1\linewidth]{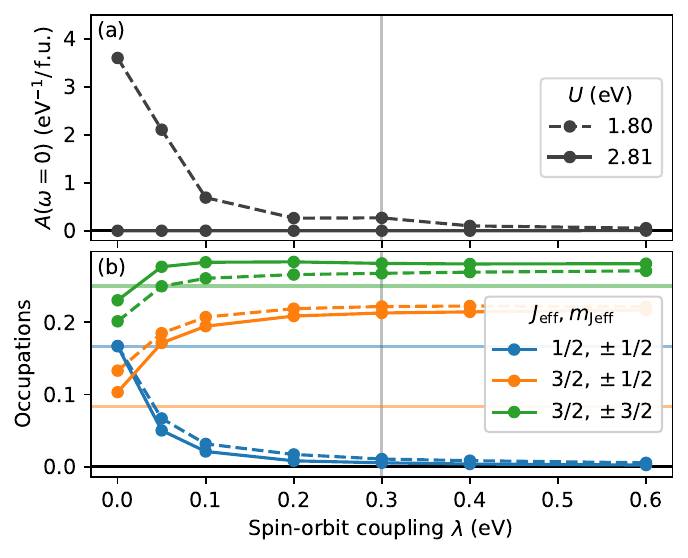}
    \caption{(a) Spectral density and (b) occupations in the \Jeff basis for $\curlyU = \SIlist{1.80;2.81}{eV}$ as a function of the SOC strength $\lambda$ obtained for the tetragonal strcuture.
    The vertical line corresponds to the realistic value of $\lambda=\SI{0.30}{eV}$, and the horizontal lines correspond to the occupations when only the $xz$ orbital is occupied. 
    }
    \label{fig:res_scan_lambda}
\end{figure}

Looking at the spectral weight $A(\omega=0)$ as a function of $\lambda$ in \pref{fig:res_scan_lambda}(a), we find that, 
also in the tetragonal phase, 
the SOC can indeed turn \BMRO from metallic to insulating, but only in the regime very close to the Mott transition, e.g., for $\curlyU = \SI{1.80}{eV}$. For $\curlyU=\SI{2.81}{eV}$, the material is always insulating, even in the complete absence of SOC, indicating that SOC does not play a crucial role in stabilizing the insulating state in this case. This is similar to what has been found for the generic three-orbital Hubbard model \cite{kim_j_2017} and for the closely related $5d^1$ double perovskite \BNOO \cite{fiore_mosca_mott_2023}.

The corresponding occupations in the \Jeff basis, shown in \pref{fig:res_scan_lambda}(b), show very similar behavior for the two different $\curlyU$ values.
For strong SOC, $\lambda \geq \SI{.30}{eV}$, the occupations are essentially constant without any significant changes compared to the case with $\lambda=\SI{0.3}{eV}$ discussed in \pref{sec:res_scan_u}, with an almost empty $\Jeff = 1/2$ doublet and a polarized $\Jeff = 3/2$ quadruplet.
Upon reducing the SOC strength, more occupation is transferred into the $\Jeff = 1/2$ states, and the polarization within the $\Jeff = 3/2$ quadruplet slightly increases. In the absence of SOC, the results, especially in the insulating case for $\curlyU = \SI{2.81}{eV}$, are relatively close to only the $xz$ orbital being occupied.

\subsection{Spectral functions and occupations}
\label{sec:analyis}

Next, we provide further analysis of the spectral features and the orbital charge distribution for the cubic and tetragonal phase for the most realistic parameter values, i.e., $\lambda = \SI{.3}{eV}$ and the cRPA interaction parameters $\curlyU = \SI{2.81}{eV}$ and $\curlyJ = \SI{.27}{eV}$. We also compare our DFT+DMFT results to the DFT+$U$ calculations of the tetragonal, paramagnetic phase from \onlinearia. 
As stated in \pref{sec:introduction}, in these DFT+$U$ calculations, the paramagnetic state was emulated by using a large supercell with random magnetic moments on the Re atoms and zero net magnetization~\cite{alling_effect_2010, trimarchi_polymorphous_2018}, which can be viewed as a static snapshot of the system in the paramagnetic state.

\subsubsection{Spectral functions}

\begin{figure}
    \centering
    \includegraphics[width=1\linewidth]{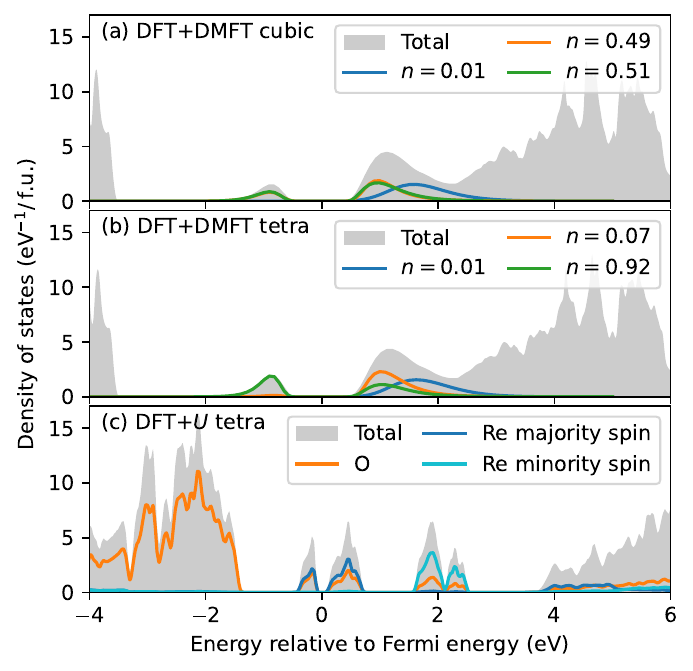}
    \caption{Spectral function from DMFT for the (a) cubic and (b) tetragonal phase, with the contribution from the frontier orbitals decomposed into the eigenstates of the local occupation labeled by their occupation. (c) Density of states of the tetragonal, paramagnetic DFT+$U$ calculation from Ref.~\cite{mansouri_tehrani_untangling_2021}, including the projection onto atomic-like Re states split into a local majority and minority channel.}
    \label{fig:pdos_para_dft}
\end{figure}

We first discuss the spectral function $A(\omega)$ obtained for the cubic phase, shown in \pref{fig:pdos_para_dft}(a), which has a large gap of around \SI{1.0}{eV}. We note that this gap is much larger than the activation energy of \SI{.17}{eV} obtained from the Arrhenius fit to the electrical resistivity \cite{hirai_successive_2019}. However, it is probably unclear whether this activation energy reflects the electronic gap of the pristine system.
To further analyze the character of the gap, we decompose the contribution from the frontier orbitals to the spectral function into the eigenstates of the local occupation. These correspond to the unoccupied $\Jeff = 1/2$ and the quarter-filled $\Jeff=3/2$ states. One can see that the lower and upper boundary of the gap is defined by the degenerate $\Jeff=3/2$ states, while the $\Jeff = 1/2$ states only contribute to higher energies in the upper Hubbard band, very similar to the DFT+DMFT calculations on \BNOO from Ref.~\onlinecite{fiore_mosca_mott_2023}.

To align the correlated states relative to the uncorrelated bands in our one-shot DFT+DMFT calculations, we subtract a double counting correction $\Sigma_\text{dc}$ from the DMFT chemical potential, using the specific form \mbox{$\Sigma_\text{dc} = (\curlyU - 2\curlyJ) (n - 1/2) = \SI{1.14}{eV}$} \cite{held_electronic_2007}, where $n=1$ is the total occupation on the Re sites.
We note that the exact form of the double counting correction is ill-defined and thus the resulting alignment is only approximate. 
Based on this, the uncorrelated states lie below \SI{-3}{eV} and above \SI{2}{eV}. Thus, while the correlated states move slightly upward relative to the uncorrelated states compared to the plain DFT bands shown in \pref{fig:band_struct}, the uncorrelated states are still far away from the gap and therefore do not play a role for the low-energy behavior of the system.

In the tetragonal phase, there is almost no change in the total spectral function compared to the cubic case, and the band gap remains \SI{1.0}{eV}, as depicted in \pref{fig:pdos_para_dft}(b).
However, the character of the states around the band gap changes drastically. There is now only one pair of eigenstates that are partially occupied, i.e., half-filled, while the other states are essentially unoccupied. This indicates that the Jahn-Teller distortion has lifted the degeneracy within the $\Jeff=3/2$ quadruplet (except for the remaining Kramer's degeneracy), while also creating some degree of mixing with the $\Jeff=1/2$ states. 

Next, we compare the DFT+DMFT spectral function for the tetragonal structure with the corresponding density of states from the paramagnetic DFT+$U$ calculations of \onlinearia.
To reduce finite-size effects, \onlinearia used five configurations with the same ionic positions in the supercell but different random magnetic moments on the Re atoms. The data depicted in \pref{fig:pdos_para_dft}(c) corresponds to the average over these five configurations.
We find very different features compared to the DFT+DMFT spectral function. The center bands are split into four distinct peaks, grouped into pairs of two separated by a gap of about \SI{2}{eV}. The Fermi energy splits the lower pair of peaks by a ratio of 1:2 in terms of their spectral weights, which results in a band gap of \SI{.2}{eV}.
Note that in the DFT+$U$ study of \onlinearia, the effective Hubbard interaction was specifically chosen to match the gap to the experimental activation energy \cite{hirai_successive_2019}.

To analyze the nature of the various peaks in the DFT+$U$ spectral function, we project the density of states from \onlinearia onto atomic-like functions in orbital and spin space, and then decompose the density of states in a local majority- and minority-spin contribution by summing up the spin projections on the local magnetic moment on a given site for all orbital characters. We then average over all Re atoms and all independent configurations.

As shown in \pref{fig:pdos_para_dft}(c), this reveals that the two lower peaks immediately below and above the Fermi level represent the local majority spin component, and the two upper peaks around \SI{2}{eV} represent the local minority spin component of the Re-\ttg states. This local spin splitting roughly coincides with the effective Hubbard interaction of \SI{1.8}{eV} used in \onlinearia. 
The band gap around the Fermi energy therefore arises from the splitting of the local majority-spin peaks, which is most likely due to the symmetry breaking by the Jahn-Teller distortion, the local spin moment, and the strong SOC.
Compared to the DFT+DMFT spectral function, the uncorrelated states lie higher in energy relative to the Re-$d$ dominated bands around the Fermi level, by about \SI{2}{eV}. However, also in DFT+$U$, this alignment depends on the ill-defined double-counting correction \cite{dudarev_electron-energy-loss_1998}.

Note that a more systematic quantitative comparison between the paramagnetic supercell DFT+$U$ calculation and our DFT+DMFT calculations is difficult since the explicit treatment of the local electron-electron interaction in the two approaches is applied to different orbitals and using different parametrizations of the interaction strength. Nonetheless, it is remarkable to see such significant qualitative differences in the corresponding spectral functions.

\subsubsection{Charge distribution in atom-like orbitals}

So far, we have employed frontier orbitals constructed from the bands immediately around the Fermi level, which are suitable for a low-energy description of \BMRO, directly map on the nominal $d^1$ charge state of the Re cation, and thus provide a clear and intuitive physical picture.
Now, to allow for a better comparison with the DFT+$U$ results of Ref.~\cite{mansouri_tehrani_untangling_2021}, we complement this with an analysis in terms of a more localized atomic-orbital-like Wannier basis, constructed from a wider energy window. Such a description in terms of localized atomic-like orbitals also allows us to analyze the degree of hybridization and quantify the local quadrupoles. Here, the atomic-like basis might be more suitable for a quantitive comparison to X-ray absorption and other experimental probes sensitive to the very local environment around the atoms.

\begin{figure}
    \centering
    \includegraphics[width=1\linewidth]{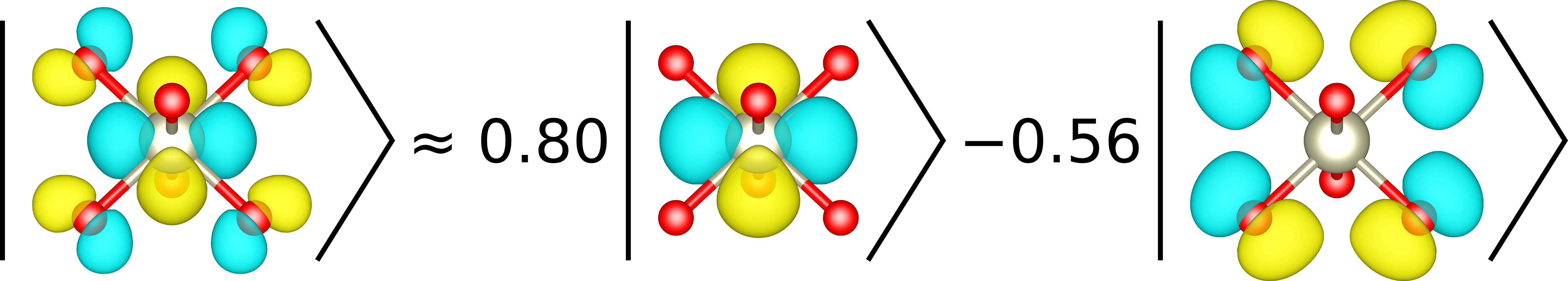}
    \caption{Wannier functions from (left) frontier orbitals and (right) localized Re-$d$ and O-$p$ orbitals. All functions have \ttg-type symmetry around the center Re atom. }
    \label{fig:bmro_projectors}
\end{figure}

These localized Wannier orbitals are constructed from all bands in the energy range between \SIlist{-13.2;7.9}{eV} relative to the Fermi energy, including all O-$2p$ and Re-$5d$ dominated bands, where the Re-$5\eg$ dominated bands need to be disentangled from higher-lying bands.
Per formula unit, we obtain five $d$-like orbitals centered on the Re sites, denoted by $\ket{\mathrm{l}_{\mathrm{Re}, m}}$, and 18 $p$-like orbitals centered on the O sites. To analyze the hybridization between these orbitals, we then proceed to combine the O-$p$ orbitals into symmetry-adapted linear combinations, $\ket{\mathrm{l}_{\mathrm{O},m}}$, which have a clearly defined symmetry with respect to the Re sites. Five of these linear combinations have a $d$-type symmetry and therefore hybridize with the Re-$5d$ orbitals. For example, a linear combination with \ttg symmetry of the O-$p$ states is shown on the right of \pref{fig:bmro_projectors}.

First, we compute the overlap between a localized orbital $\ket{\mathrm{l}_{Cm}}$, where $C$ is the atomic character of the localized state and can thus be Re or O, and a frontier orbital $\ket{\mathrm{f}_{m'}}$ on the same \ReOoct octahedron.
Using the fact that the Kohn-Sham states $\ket{\bm k i}$ at reciprocal vector $\bm k$ with band index $i$ form a complete ortho-normal basis inside the chosen energy window, we obtain:
\begin{align}
    \braket{\mathrm{l}_{Cm}}{\mathrm{f}_{m'}} = \sum_{\bm k i} \braket{\mathrm{l}_{Cm}}{\bm k i} \braket{\bm k i}{\mathrm{f}_{m'}} ,
    \label{eq:overlap_bases}
\end{align}
where the overlaps between Wannier functions and Kohn-Sham states are given by the $U(\bm k)$ and $U^\mathrm{dis}(\bm k)$ matrices obtained from Wannier90 \cite{mostofi_wannier90_2008}. We note that, due to the gauge freedom of the Bloch functions, these matrices crucially need to come from the same Kohn-Sham states at the same DFT iteration for both types of Wannier orbitals.

We find $\alpha \equiv \braket{\mathrm{l}_{\mathrm{Re},m}}{\mathrm{f}_m} = 0.80$ and $\beta \equiv \braket{\mathrm{l}_{\mathrm{O},m}}{\mathrm{f}_m} = -0.56$ for all three orbitals with \ttg symmetry in both the cubic and tetragonal phase. By symmetry, there is no overlap between orbitals with different orbital characters, $m \neq m'$.
The resulting decomposition of the frontier orbital is schematically visualized in \pref{fig:bmro_projectors}.
Thus, the frontier orbitals consist to $\alpha^2 = \SI{64}{\percent}$ of Re-$\ttg$ orbitals and $\beta^2 = \SI{31}{\percent}$ of O-$p$ orbitals, with the missing \SI{4}{\percent} belonging to other orbitals that we do not explicitly take into account. 
Such strong hybridization can be expected from the relatively delocalized nature of the Re-$5d$ states and is comparable to results from a cluster model for \BNOO fitted to experimental data, which reported a hybridization of \SI{50}{\percent} Os and \SI{50}{\percent} O \cite{erickson_ferromagnetism_2007}.

We can now compute the occupation in this localized basis from:
\begin{align}
    \bra{\mathrm{l}_{Cm}} n \ket{\mathrm{l}_{C'm'}} =&
    \sum_{\mu\mu'} \braket{\mathrm{l}_{Cm}}{\mathrm{f}_\mu}
    \bra{\mathrm{f}_\mu} n \ket{\mathrm{f}_{\mu'}}
    \braket{\mathrm{f}_{\mu'}}{\mathrm{l}_{C'm'}} \nonumber\\
    &+ \sum_{\substack{\bm k i \\ \epsilon_{\bm k i} < \bar\epsilon}}
    \braket{\mathrm{l}_{Cm}}{\bm k i}
    \braket{\bm k i}{\mathrm{l}_{C'm'}} .
\end{align}
The first term projects the occupations $\bra{\mathrm{f}_\mu} n \ket{\mathrm{f}_{\mu'}}$ of the central bands around the Fermi energy from the DMFT calculation onto the localized orbitals, whereas the second term captures the fully occupied bands below these correlated bands by summing over all bands with energies $\epsilon_{\bm k i}$ below $\bar\epsilon = \SI{-1}{eV}$ [\textit{cf.} \pref{fig:band_struct}(a)].

The strong hybridization with the oxygen leads to 2.7 electrons in the atomic-like Re-\ttg and 1.2 electrons in the corresponding \eg orbitals and therefore a total of 3.8 $d$ electrons on the Re. This differs significantly from the nominal $d^1$ occupation in the frontier basis and is quite similar to the 4.0 $d$ electrons found in the paramagnetic DFT+$U$ calculations of \BMRO \cite{mansouri_tehrani_untangling_2021}. It is also comparable to the occupation of up to 6.3 electrons reported in an atomic-like basis for the nominal-$d^2$ double perovskites Ba$_2$\textit{M}OsO$_6$~\cite{fiore_mosca_modeling_2022}.

We now decompose the occupation matrix in the frontier and localized basis sets into multipoles as defined in Ref.~\onlinecite{bultmark_multipole_2009} and implemented in Ref.~\onlinecite{merkel_multipyles_2023}. Here, we focus on the SOC moment $\langle \bm l \cdot \bm s \rangle$, which quantifies the effect of the SOC in \pref{eq:h_soc} on the spin and charge density, and on the pure charge quadrupoles centered on the Re sites. We compute the multipoles in the localized basis from the density matrices $\bra{\mathrm{l}_{Cm}} n \ket{\mathrm{l}_{C'm'}}$ for the different $C$ and $C'$, and thus obtain a Re-Re, a Re-O, and an O-O contribution (which all contribute to the multipoles centered around the Re sites). 
Note that the multipoles in frontier and localized basis are related by the approximate decomposition of the frontier into localized orbitals depicted in \pref{fig:bmro_projectors}. Since the multipole decomposition is a linear operation on the matrix elements of the density matrix, the considered frontier multipoles $w$ are related to the corresponding localized ones by
\begin{align}
    w_\text{frontier} \approx \alpha^2 w_\text{Re-Re} + 2 \alpha\beta w_\text{Re-O} + \beta^2 w_\text{O-O} .
    \label{eq:multipole_weighting}
\end{align}
Here, we assume that the bands below $\bar\epsilon$ do not contribute to the frontier multipoles and use the fact that all overlaps and multipoles are real.

\begin{table}
    \centering
    \caption{Multipoles corresponding to the effect of SOC and charge quadrupoles with \eg symmetry centered on the Re sites, obtained in the frontier and localized basis sets. The $\pm$ signs indicate the two equivalent Re sites.
    Note that in the normalization used here \cite{bultmark_multipole_2009}, a negative $z^2$ quadrupole corresponds to excess charge along the $z$ direction.
    }
    \label{tab:multipoles_bases}
    \begin{ruledtabular}
    \begin{tabular}{lrrrr}
        & frontier & \multicolumn{3}{c}{localized} \\
        & & Re-Re & Re-O & O-O \\
        \cline{1-1} \cline{2-2} \cline{3-5}
        \rule{0pt}{1\normalbaselineskip}%
        cubic $\langle \bm l \cdot \bm s \rangle$ & $-0.48$ & $-0.31$ & $0.22$ & $-0.15$ \\
        tetra $\langle \bm l \cdot \bm s \rangle$ & $-0.48$ & $-0.31$ & $0.22$ & $-0.15$ \\
        tetra $z^2$ & $-0.13$ & $-0.08$ & $0.06$ & $-0.04$ \\
        tetra $x^2-y^2$ & $\mp0.50$ & $\mp0.32$ & $\pm0.22$ & $\mp0.15$ \\
    \end{tabular}
    \end{ruledtabular}
\end{table}

The resulting multipoles are shown in \pref{tab:multipoles_bases}. Note that the O-O and Re-O contributions in the localized basis still correspond to the multipole centered at the Re site. 
The SOC moment $\langle \bm l \cdot \bm s \rangle$ is close to $-1/2$ in the frontier basis in both the cubic and tetragonal phase, which is equivalent to the single electron residing in the $\Jeff=3/2$ states discussed in the context of, e.g., \pref{fig:res_scan_u}. In the localized basis, we can see that the effect of the SOC distributes over Re and O. Therefore, even though the SOC is a local effect caused mostly by the Re atoms, due to hybridization, the occupations of the O are also affected.

In cubic \BMRO, the quadrupoles on the Re sites are exactly zero since the charge density is always cubic in our calculations, as verified for a larger range of temperatures in \pref{app:res_scan_temp}.
In the tetragonal phase, the $x^2-y^2$ and $z^2$ quadrupoles appear, which are depicted in \pref{fig:crystal_struct}(b, c) and are the only quadrupoles allowed in \Pftmnm symmetry.
Similar to the SOC moment, we find that the local quadrupole is close to saturation in the frontier basis, with a large $x^2-y^2$ component. In the localized basis, again both Re and O play a role, with the dominant contributions coming from the Re-Re and Re-O terms, which enter the total quadrupole moment weighted by $\alpha^2 = 0.64$ and $2\alpha\beta = -0.90$, respectively, as discussed in \pref{eq:multipole_weighting}.
Since the quadrupole moments are essentially just a specific way to quantify the local orbital polarization, this shows that an orbital polarization can also be found on the ligands, and even more pronounced in the mixed component representing the hybridization between transition-metal ion and ligand. 

\begin{figure}
    \centering
    \includegraphics[width=1\linewidth]{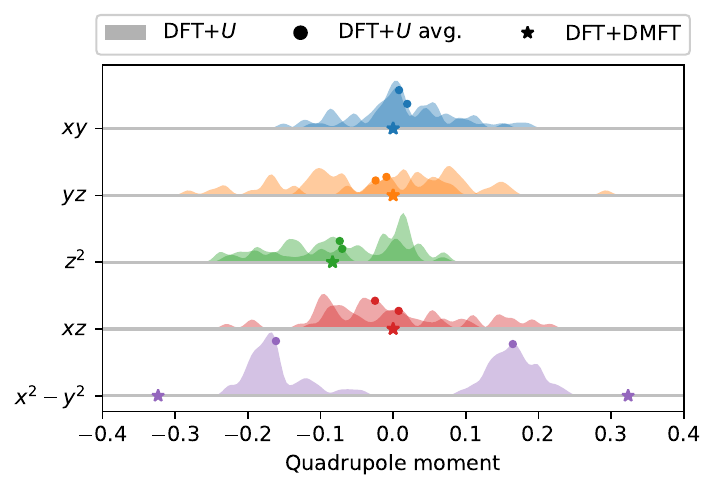}
    \caption{Charge quadrupoles on the Re sites in tetragonal \BMRO, evaluated in a direct comparison between the DFT+DMFT calculations and the distributions of quadrupoles on the two different Re sites from the paramagnetic-supercell DFT+$U$ calculations of \onlinearia.}
    \label{fig:quadrupoles_comp_dftu}
\end{figure}

For the tetragonal phase, we can now directly compare the Re-Re contribution of the quadrupoles%
    \footnote{Note that in \onlinearia a different coordinate system was used, where the $x$ and $y$ axes are rotated by \ang{45} about the $z$ axis away from the Re-O bonds. We therefore transform the quadrupoles from \onlinearia to our choice of axes, i.e. oriented along the Re-O bonds. This swaps $xy$ and $x^2-y^2$ and linearly combines $xz$ and $yz$.}
in the localized basis to the data from \onlinearia.
Due to the random magnetic moments in the paramagnetic supercell used in DFT+$U$, all Re atoms become symmetry-inequivalent, and in \pref{fig:quadrupoles_comp_dftu} we thus show the resulting distribution of quadrupoles (over all five independent configurations considered in \onlinearia). We show separate distributions for the two Re sublattices with the $Q_2$-related octahedral elongation along $x$ or $y$ direction, respectively. Then, we compare the average of these two distributions to the quadrupoles from DFT+DMFT on the corresponding sites. 

In the DFT+DMFT calculations, the quadrupoles with \ttg symmetry are zero, as mandated by \Pftmnm symmetry. This agrees with the DFT+$U$ calculations, which show a wide distribution, but with an average always absolutely smaller than 0.03, so that also from DFT+$U$, \onlinearia concluded that these quadrupoles are zero.
The $x^2-y^2$ quadrupole clearly orders antiferroically in both DFT+DMFT and DFT+$U$, as expected from the symmetry between the different distortive directions.
As discussed before, a quantitative comparison between the DFT+DMFT and DFT+$U$ results is difficult, due to the different treatment of the local electron-electron interaction, which can potentially explain the approximate factor of two in difference between the predicted $x^2-y^2$ quadrupoles from the two methods. 
Finally, DFT+DMFT confirms the ferroic ordering of the $z^2$ quadrupoles, which was not completely obvious from \onlinearia due to the rather broad distribution.

\section{Conclusions and Summary}

We have used DFT+DMFT calculations to establish that \BMRO is Mott-insulating both in the cubic and tetragonal paramagnetic phases. In both cases, the insulating state can be classified as a regular Mott insulator, originating from the Coulomb repulsion $\curlyU$, with a critical $\curlyU$ for the metal-insulator transition much smaller than the value of \SI{2.8}{eV} we calculated with cRPA. Even though the spin-orbit coupling is strong, leading to a splitting of around \SI{.45}{eV} between $\Jeff=3/2$ and $\Jeff=1/2$ within the low-energy states, it does not significantly influence the metal-insulator transition. This agrees with the expected behavior for $d^1$ systems \cite{kim_j_2017, fiore_mosca_mott_2023}.

The spectral function of \BMRO around the Fermi energy is dominated by two Hubbard peaks, which form a gap of about \SI{1.0}{eV} in both cubic and tetragonal phases. However, the orbital character of the Hubbard peaks changes across the cubic-tetragonal transition, in line with the corresponding change in the local Re occupations in the frontier orbital basis. The insulating, cubic phase can be accurately described in a nominal $d^1$ picture with a quarter-filling of the $\Jeff=3/2$ states due to the spin-orbit coupling. In the tetragonal, paramagnetic phase, we find approximate half-filling of a state determined by the combined effects of the spin-orbit coupling and the tetragonal crystal field, with a remaining two-fold degeneracy due to time-reversal symmetry.

We also analyze the local electron distribution around the Re sites in terms of a more localized atomic-orbital-like basis. These orbitals might be more suitable for comparison with experimental probes that are sensitive to the immediate atomic environment such as resonant X-rays, and also allow for a better comparison with the DFT+$U$ results of \onlinearia.  
Two complementary but equivalent pictures emerge when postprocessing the charge and spin density with either frontier or atomic-like Wannier orbitals. The former directly maps on the nominal $d^1$ picture, with physics comparable to model studies \cite{kim_j_2017, triebl_spin-orbit_2018, iwahara_vibronic_2023}, while the latter corresponds to an atomic picture, where both Re and O orbitals are influenced by the tetragonal distortion and the spin-orbit coupling. Notably, even though spin-orbit coupling is a local effect coming mostly from the heavy Re atoms, its signature in the hybridized bands around the Fermi energy also strongly affects the occupation of the O orbitals, indicated by a nonzero component of the spin-orbit coupling $\langle \bm l \cdot \bm s \rangle$ moment.

Finally, we compare our DFT+DMFT results to DFT+$U$ calculations for the tetragonal phase of \BMRO, which have used large supercells with random magnetic moments to emulate the paramagnetic state~\cite{mansouri_tehrani_untangling_2021}. 
We find good qualitative agreement between the two methods for the quadrupole moments around the Re sites. Our calculations even corroborate the existence of a ferroically ordered $z^2$ quadrupole component, which was slightly obscured in the DFT+$U$ calculation of \onlinearia by the fluctuations related to the random magnetic moments but has also been reported experimentally~\cite{hirai_successive_2019, hirai_detection_2020, soh_spectroscopic_2023}. 
In contrast, we find noticeable differences in the spectral functions obtained within DFT+DMFT and DFT+$U$, where in the latter case, the Re-dominated density of states around the Fermi level splits into four peaks with well-defined local spin characters, which is rather different from the DFT+DMFT spectral function.
Part of this difference can potentially be explained by the differences in how the local electron-electron interaction is treated in the two cases. Nevertheless, this comparison raises the question of how comparable the spectral features of paramagnetic DFT+$U$ and DFT+DMFT calculations are, especially in the presence of spin-orbit coupling where the orbital occupation is strongly influenced by local magnetic moments.

\begin{acknowledgments}
We thank Sophie Beck for implementing the real-valued formulation of SOC in solid\_dmft and Andrea Urru for the discussions and help in implementing the multipole decomposition in Python.
This research was supported by ETH Zurich and a grant from the Swiss National Supercomputing Centre (CSCS) under project ID s1128.
\end{acknowledgments}

\appendix

\section{Effect of temperature}
\label{app:res_scan_temp}

The critical temperature of \SI{33}{K} for the transition from cubic to tetragonal is much lower than the electronic temperature of \SI{290}{K} used in our DMFT calculations. To verify that this high electronic temperature does not crucially affect the results obtained in our DFT+DFMT calculations for the tetragonal phase, we now present additional calculations where we vary the electronic temperature and monitor the effect on the orbital occupations. We also test whether it is possible to stabilize spontaneous orbital order in the cubic phase, or whether the charge density always preserves the \Fmtm symmetry. Here, we again use the realistic parameters of $\lambda = \SI{.3}{eV}$, $\curlyU = \SI{2.81}{eV}$, and $\curlyJ = \SI{.27}{eV}$.

\begin{figure}[b]
    \centering
    \includegraphics[width=1\linewidth]{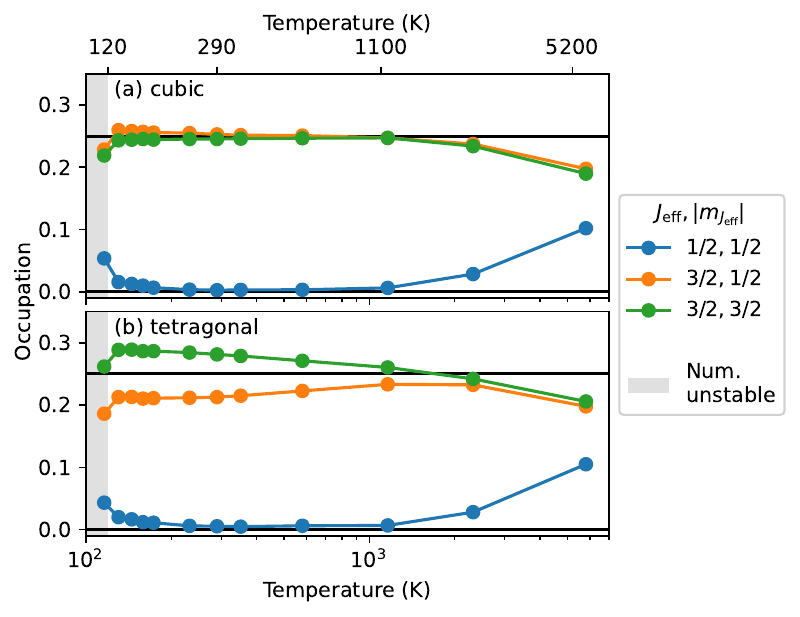}
    \caption{Occupations in the \Jeff basis for the cubic and tetragonal phase as a function of the electronic temperature in the DMFT calculation. The upper temperature axis indicates the points corresponding to the boundary of the numerical instability region (\SI{170}{K}), room temperature (\SI{290}{K}), and the thermal energies equal to the tetragonal crystal-field splitting (\SI{1100}{K}) and the SOC splitting (\SI{5200}{K}).
    The region of numerical instability is defined when the average sign in the CT-HYB solver drops below $0.1$.
    }
    \label{fig:res_scan_temp}
\end{figure}

\pref{fig:res_scan_temp} shows the occupation as a function of electronic temperature, both for the cubic and tetragonal case. In the tetragonal case, the polarization of the $\Jeff=3/2$ quadruplet depends only weakly on the temperature in the range from \SIrange{120}{290}{K}. Below \SI{120}{K}, indicated by the gray shaded region in \pref{fig:res_scan_temp}, the calculations become numerically unstable due to the fermionic sign problem appearing in the quantum Monte-Carlo solver. Nevertheless, it seems reasonable to assume that, for all temperatures below \SI{290}{K}, the polarization of paramagnetic \BMRO is already more or less saturated. This \emph{a posteriori} justifies the choice of electronic temperature in the DFT+DMFT calculations discussed in the previous sections. 

Note that at higher temperatures, thermal effects first reduce the polarization within the $\Jeff = 3/2$ states, and then the polarization between $\Jeff = 3/2$ and $\Jeff = 1/2$ states, at temperatures of around \SIlist{1100;5200}{K}, where the thermal energy is roughly equivalent to the respective energy splittings.

In cubic \BMRO, for all temperatures above \SI{120}{K} which are accessible with our DFT+DMFT calculations, we find no tendency towards spontaneous polarization, in agreement with the low experimental transition temperature.
At higher temperatures, we find the suppression of SOC-induced polarization, as already observed in the tetragonal phase.

\bibliography{bibfile}

\end{document}